\documentstyle[aps,epsfig,multicol]{revtex}
\def\breakon{\end{multicols}\widetext\vspace{-.2cm}
\noindent\rule{.48\linewidth}{.3mm}\rule{.3mm}{.3cm}\vspace{.0cm}}

\def\breakoff{\vspace{-.2cm}
\noindent
\rule{.52\linewidth}{.0mm}\rule[-.27cm]{.3mm}{.3cm}\rule{.48\linewidth}{.3mm}
\vspace{-.3cm}
\begin{multicols}{2}
\narrowtext}

\newcommand{\be}{\begin{equation}}
\newcommand{\ee}{\end{equation}}
\newcommand{\bea}{\begin{eqnarray}}
\newcommand{\eea}{\end{eqnarray}}
\newcommand{\HH}{{\cal H}}

\newcommand{\tr}{{\rm tr\/}\,}

\newcommand{\la}{\langle}
\newcommand{\ra}{\rangle}

\newcommand{\lp}{\left(}
\newcommand{\rp}{\right)}
\renewcommand{\phi}{\varphi}

\begin{document}

\title{Tunable Fermi-Edge Resonance in an Open Quantum Dot}
\author{D. A. Abanin and L. S. Levitov}
\address{
Department of Physics,
Center for Materials Sciences \& Engineering,
Massachusetts Institute of Technology, \\
77 Massachusetts Ave.,
Cambridge, MA 02139}

\maketitle
\begin{abstract}
Resonant tunneling in an open mesoscopic quantum dot is proposed 
as a vehicle 
to realize a tunable Fermi-edge resonance 
with variable coupling strength.
We solve the x-ray edge problem for 
a generic nonseparable scatterer and apply it to 
describe tunneling in a quantum dot.
The tunneling current power law exponent 
is linked to the S-matrix of the dot.
The control of scattering 
by varying the dot shape and coupling to the leads
allows to explore a wide range 
of exponents. 
Transport 
properties, such as weak localization,
mesoscopic conductance fluctuations, and sensitivity 
to Wigner-Dyson ensemble type,
have their replicas in
the Fermi-edge singularity.
\end{abstract}


\vspace{-5mm}
\begin{multicols}{2}

\narrowtext

Quantum dots host a number 
of interesting quantum transport phenomena, 
such as Coulomb blockade\cite{FultonDolan,Kastner}, Kondo effect\cite{Kondo}
weak localization and universal conductance fluctuations\cite{QDfluctuations}. 
Electron scattering inside
the dot as well as the dot-lead coupling can be controlled
externally by gates,
which makes it possible 
to reveal phenomena of interest
by varying system parameters.
Such tunability has been exploited
to demonstrate\cite{Pustilnik,Sasaki00,Jeong01,Oreg03}
new exotic varieties of Kondo effect.
In this article we 
propose to employ mesoscopic dots,
in a similar controllable fashion,
to study the Fermi-edge singularity (FES).

FES is a fundamental manifestation of many-body 
physics taking place when an electron
with energy just above the Fermi level tunnels into a metal,
while leaving a localized hole behind.
After tunneling, the electron forms 
a quasiresonance due to interaction with the hole.
This strongly affects the transition rate which is typically 
found to be a power law function of Mahan-Nozieres-deDominicis form 
$A(\epsilon)\propto (\epsilon-\epsilon_F)^{-\alpha}$. 
Similar to the Kondo problem,
the FES problem~\cite{Mahan,Nozieres_deDominicis} is one of few 
many-body problems exactly solvable
in the nonperturbative regime of strong interaction.

First discovered in the 60's in the context of x-ray absorption 
in metals~\cite{Mahan,Nozieres_deDominicis}, 
the FES physics has found many other applications. 
In 1992, Matveev and Larkin~\cite{MatveevLarkin} 
considered resonant tunneling and
predicted 
a power law singularity, identical to FES,
as a function of the resonance 
position relative to the Fermi level. 
In this case, the exponent $\alpha$ 
in the tunneling $I-V$ characteristic 
is controlled by
interaction of 
the tunneling electron and localized hole. 
The latter is system-specific, and depends on scattering phases
and screening via Friedel sum rule.

Below we generalize the theory~\cite{MatveevLarkin} to describe resonant
tunneling into an open quantum dot.
Chaotic scattering in the dot,
returns the tunneling electron many times
to the hole, which enhances the FES singularity
and makes it `tunable', i.e. scattering-dependent.
(In a noninteracting mesoscopic system\cite{LernerRaikh}, 
multiple returns to a resonant level
are known to produce weak localization and conductance fluctuations.)
While charging effects
may interfere with resonant 
tunneling\cite{Gramespacher,Bascones}, in open dots one 
can ignore charge fluctuations and
focus on the interplay of scattering and interaction
with localized hole which forms FES.

Manifestations of FES have 
been observed in tunnel junctions~\cite{Geim}
and 
in low temperature 
telegraph noise~\cite{Cobden}. The role of scattering 
in quantum dots can 
be studied by resonant tunneling spectroscopy~\cite{Marcus}.

The canonical theory~\cite{Nozieres_deDominicis},
based on separable scatterer model, is
difficult to adapt to mesoscopic scattering.
The crucial problem  
arises from noncommutativity of the S-matrices before and after
electron release in the dot, rendering the separable model,
along with the bosonization approach\cite{SchotteSchotte} used to handle it, 
irrelevant. 
Our approach builds
on the Yamada and Yosida theory~\cite{YamadaYosida} 
of Anderson orthogonality for
multichannel nonseparable scatterer, recently advanced
by Muzykantskii \emph{et al.}~\cite{Muzykantskii_Adamov,Muzykantskii_Xray},
as well as on Matveev phase shift approach~\cite{Matveev} to charge 
fluctuations in quantum dots.

The theory presented below yields an exact relation
of the FES exponent
with the quantum dot S-matrix and reveals that the exponent
structure is similar to that in the separable scatterer case.
The orthogonality catastrophe due to Fermi sea shakeup 
by switching of charge state at tunneling accounts only for 
one, negative part of the FES exponent, while the  
leading, positive part arises from interaction
in the final state. 
Applying the result to the quantum dot problem, 
we find that by varying
the dot scattering parameters the exponent $\alpha$ can be tuned 
to any value in the weak or strong coupling regime. 

Our results for open dots complement the work on the orthogonality
catastrophe~\cite{AleinerMatveev,Smolyarenko,Gefen} and FES~\cite{Hentschel} 
in closed quantum dots which use the exact one particle states and energies
to express the many-body overlap and transition rate. 
The enhancement of orthogonality by disorder, discussed by
Gefen \emph{et al.}~\cite{Gefen}, has the same underlying physics as
our backscattering-enhanced FES.

\begin{figure}[t]
\includegraphics[width=3.0in]{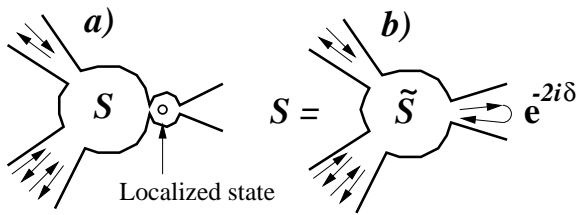}
\vspace{0.1cm}
\caption[]{
a) An open quantum dot weakly coupled to a small closed dot which 
holds a localized electron that can tunnel into the open dot.
b) Relation of the open dot S-matrix $S$ and an auxiliary
matrix $\tilde S$ is illustrated.
The latter describes the dot with an extra open channel added
to incorporate backscattering on the small dot charge state.
}
\label{fig1}
\end{figure}

The geometry of interest is pictured in Fig.~\ref{fig1} (a). 
We consider tunneling from a small dot
which holds few electrons and 
has a large charging energy, into an open mesoscopic dot. 
The latter is characterized by a $N\times N$ S-matrix, 
where $N$ is the number of channels 
connecting the dot and the leads. 
The interaction of electrons in the 
open dot 
with the localized hole in the small dot is described\cite{Matveev} by 
the backscattering phase $\delta$ 
in the channel connecting the two dots (Fig.~\ref{fig1} (b)). 

How does the dot S-matrix depend on backscattering? 
The answer is found most easily by considering  
an auxiliary scattering
problem with an additional channel which describes 
the point contact between the dots. This defines 
an extended S-matrix $\tilde S$ of size $(N+1)\times (N+1)$. 
The physical matrix $S$ can be linked with
the auxiliary matrix $\tilde S$ 
by imposing the quasiperidicity relation 
$u^{(out)}_{N+1}=e^{-2i\delta}u^{(in)}_{N+1}$ on the in and out
components of the added channel,
and eliminating these components from
the scattering relation $u^{(out)}_{i}=\tilde S_{ij}u^{(in)}_{j}$.
We obtain 
\be\label{eq:SS}
S_{ij}=\tilde S_{ij}+\frac{\tilde S_{i(N+1)}\tilde S_{(N+1)j}}{e^{-2i\delta}-r}
\,,\quad i,j=1...N
\ee
with $r\equiv \tilde S_{(N+1)(N+1)}$ the backscattering amplitude 
in the extended picture. One can verify that $S$, defined by (\ref{eq:SS}),
is unitary provided that $\tilde S$ is unitary. The relation between 
$S$ and $\tilde S$ is illustrated graphically in Fig.~\ref{fig1} (b).

We emphasize that the parameters $r$ and $\delta$ which appear 
together in Eq.(\ref{eq:SS}) and below
describe different physics
and, in particular, arise on different length scales.  
The phase shift $\delta$ 
is a 
constant determined by the effects within a screening length
from localized hole.
In contrast, the quantity $r$, describing 
transport in the interior of the dot, 
is sensitive to the dot shape, and thus is tunable.

The utility of the $\delta$-dependent S-matrix (\ref{eq:SS})
can be assessed by using the result 
for the orthogonality catastrophe with 
nonseparable scatterer. In the latter problem, one is 
interested in the overlap of the many-body ground states 
with different $S=S_{0,1}$.
Yamada and Yosida~\cite{YamadaYosida}
derived a remarkably simple formula for this overlap,
\be\label{eq:YY}
\la 1|0\ra = \exp\lp - \frac{\beta^2}{2} \ln N\rp
,\quad
\beta^2=\frac1{4\pi^2} | \tr \ln^2 S_1S_0^{-1}|
\ee
with $N$ the number of particles per scattering channel.

For two different phases 
$\delta$, $\delta'$ in (\ref{eq:SS}), the 
compound matrix 
$R=S(\delta')S^{-1}(\delta)$ that appears in (\ref{eq:YY})
is equal to
\be\label{eq:R}
R_{ij}=\delta_{ij}+\lp \frac{U(\delta)}{U(\delta')}\!-1\!\rp v_i^\ast v_j
, \quad
U(\delta)=\frac{e^{-2i\delta}-r}{e^{-2i\delta}r^\ast -1}
\ee
with $v_i=\tilde S_{(N+1)i}(1-|r|^2)^{-1/2}$, $i=1...N$, the normalized 
column of $\tilde S$. Remarkably, $R$ differs from a unit matrix 
only by a matrix of rank one. It means that, 
despite the complex dependence (\ref{eq:SS})
on the scattering phase $\delta$ that appears to affect
the entire $N\times N$ matrix $S$,
the orthogonality problem is 
effectively a single channel-like. The overlap
$\la 0|1\ra$ is described by Eq.(\ref{eq:YY}) with  
\be\label{eq:beta}
\beta= \frac1{2\pi} \,{\rm Im}\, \ln \lp \frac{U(\delta)}{U(\delta')} \rp
\ee
Thus $\beta$ depends on transport
in the dot solely via the backscattering amplitude $r$. 
Both the modulus of $r$ and its phase, being functions
of the dot shape and dot-lead transmissions, are under experimental control,
and thus $\beta$ can be tuned to any value in the interval $0<\beta<1$.

We analyze the FES problem below for the scatterers 
(\ref{eq:SS}),
and find a relation between the FES 
and the orthogonality exponents identical to the single channel case,
\be\label{eq:alpha_beta}
\alpha=2\beta-\beta^2
\ee
This is not entirely unexpected, given that the above analysis 
reveals hidden single-channel character of the orthogonality problem.
However, since the canonical
FES theory~\cite{Mahan,Nozieres_deDominicis}
is limited to the separable scatterer situation, the relation
(\ref{eq:alpha_beta}) cannot be deduced directly. 
Instead, we shall 
develop an approach 
for a generic S-matrix, and then specialize to 
the quantum dot case (\ref{eq:SS}).

Turning to the analysis of the FES problem, we consider the tunneling electron 
Green's function~\cite{Mahan,Nozieres_deDominicis}
\be\label{eq:G_general_el}
G(\tau)
=\tr \lp \hat \psi(0) e^{-i\HH_1\tau} \hat \psi^+(0) e^{i\HH_0 \tau} 
\hat\rho_e\rp
\ee
with interaction included in the Hamiltonians $\HH_{1,2}$ which
describe electron scattering by the charged/uncharged 
localized state. 
Here \mbox{$\hat \psi(\tau)=\sum_\alpha u_\alpha \hat a_\alpha(\tau)$}
is the operator of a tunneling electron 
with $\hat a_\alpha$ labelled by energy and scattering channel,
while
$\hat\rho$ is electron density matrix
\be\label{eq:rho_e}
\hat\rho_e = \frac1{Z} \exp\lp -\beta\sum_\alpha \epsilon_\alpha \hat a^+_\alpha \hat a_\alpha\rp
\,,\quad
\beta^{-1}=k_{\rm B}T
\ee
with the normalization factor 
$Z=\prod _{\alpha} \lp 1+e^{-\beta\epsilon_\alpha} \rp$.
The original approach~\cite{Nozieres_deDominicis} employs
a diagrammatic expansion of 
(\ref{eq:G_general_el})
and, using the closed loop calculus, 
expresses it through electron Green's function  
describing time-dependent scattering at $0<t<\tau$.
Then the series for the Green's function are resummed in order to 
replace the scattering potential by the S-matrix. 
The resummed series, treated using 
Dyson equation
in the time domain, lead to a singular integral equation
that can be solved 
using a particular variety of the Wiener-Hopf method.

Here we proceed differently, trying to avoid the 
diagrammatic expansion altogether. This has a two-fold advantage.
Firstly, we shall be able to introduce the single particle 
S-matrices 
at an early stage of the calculation,
thereby bypassing the resummation problem.
Secondly, our approach will apply to noncommuting S-matrices.

As a first step, we use the commutation relations
\be
\hat a^+_\alpha \hat \rho_e = e^{\beta\epsilon_\alpha} \hat \rho_e \hat a^+_\alpha 
\,,\quad
\hat a^+_\alpha e^{i\HH_0\tau} = e^{-i\epsilon_\alpha \tau} e^{i\HH_0 \tau} \hat a^+_{\alpha}
\ee
to rewrite Eq.\,(\ref{eq:G_general_el}) as
\be\label{eq:G_aa'}
G(\tau)
\!=\!\sum_{\alpha,\alpha'}\! u^\ast_{\alpha'} u_\alpha
e^{\beta\epsilon_{\alpha'}}  e^{-i\epsilon_{\alpha'} \tau}
\tr \!\lp e^{-i\HH_1 \tau}\!e^{i\HH_0 \tau} \!\hat\rho_e \hat a^+_{\alpha'} \hat a_\alpha  \rp
\ee
where $\alpha$, $\alpha'$ label single particle band states.

Next, we note that
the quantities $e^{-i\HH_1 \tau}$, 
$e^{i\HH_0 \tau}$, $\hat\rho_e$ are exponentials of 
operators quadratic in $\hat a_\alpha$, $\hat a^+_{\alpha'}$, 
which allows to write 
their product as 
\be\label{eq:w_def}
e^{-i\HH_1 \tau}e^{i\HH_0 \tau} \hat\rho_e = Z^{-1} \exp\lp \sum_{\beta,\beta'}
w_{\beta\beta'}\hat a^+_\beta\hat a_{\beta'}\rp
\ee
where the operator $\hat w$, defined by Eq.\,(\ref{eq:w_def})
and to be found in an explicit form below, 
acts in the single electron Hilbert space.
With the help of the definition (\ref{eq:w_def})
the trace in Eq.\,(\ref{eq:G_aa'}) can be expressed
through $\hat w$ as follows:
$$
\tr \lp e^{-i\HH_1 \tau}e^{i\HH_0 \tau} \hat\rho_e 
\hat a^+_{\alpha'} \hat a_\alpha
\rp
= 
\frac{\det \lp \hat 1+e^{\hat w}\rp}{Z}
\lp \hat 1+e^{-\hat w}\rp^{-1}_{\alpha\alpha'} 
$$
which reduces the FES problem to analyzing the operators
$\hat 1+e^{\pm \hat w}$. 
As we find shortly,  
the latter are related to
the single-particle S-matrix and energy distribution. 
The electron statistics is thus fully accounted for by 
the algebra involved in the construction 
of the operator $\hat 1+e^{\hat w}$ and its determinant,
while the solution of the time-dependent scattering problem
amounts to computing the inverse $\lp \hat 1+e^{-\hat w}\rp^{-1}$.
The explicit separation of the many-body and 
the single-particle effects provides an efficient treatment of
the FES problem.

\begin{figure}
\includegraphics[width=3.1in]{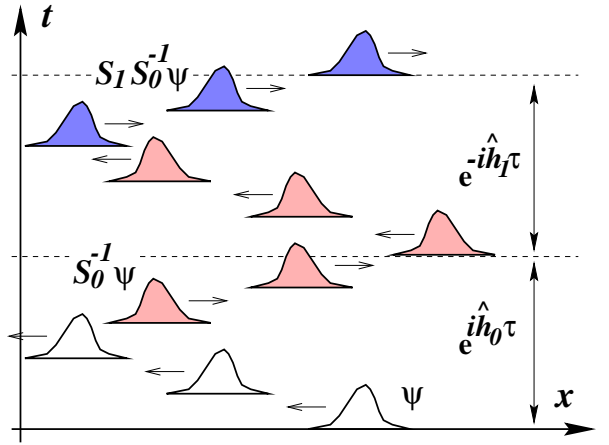}
\vspace{0.1cm}
\caption[]{
Schematic forward and backward scattering time evolution 
$e^{-i\hat h_1\tau}e^{i\hat h_0\tau}$,
with ballistic transport outside the scattering region.
The compound S-matrix 
$R=S_1S_0^{-1}$ accounts for sequential scattering described by 
the S-matrices $S_{0,1}$ corresponding to the hamiltonians $\hat h_{0,1}$.
}
\label{fig2}
\end{figure}

To make 
progress, we 
use the idea of Ref.~\cite{Klich} to
link $e^{\hat w}$ with single-particle quantities.
From Baker-Hausdorff series for
$\ln(e^Ae^B)$ in terms of multiple commutators of $A$ and $B$, 
noting the correspondence between 
the commutator algebra of the many-body operators quadratic in 
$a_{\alpha}$, $a_{\alpha'}^+$
and the single-particle operators, 
we find
\be\label{eq:expw}
e^{\hat w}=e^{-i\hat h_1 \tau}e^{i\hat h_0 \tau}e^{-\beta\hat\epsilon}
\ee
Here the operators $\hat h_{0,1}$ and $e^{-\beta\hat\epsilon}$ 
are related to the single particle 
Hamiltonian and density matrix (\ref{eq:rho_e}) as
\be
\HH_{0,1}=\sum_{\alpha\alpha'} (\hat h_{0,1})_{\alpha\alpha'}a^+_\alpha
a_{\alpha'}
\,,\quad
e^{-\beta\hat\epsilon} =
e^{-\beta\epsilon_\alpha}\delta_{\alpha\alpha'}
\ee
With the help of the result (\ref{eq:expw}), 
defining $\hat n=(1+e^{\beta\hat\epsilon})^{-1}$,
the determinant $\det\lp 1+e^{\hat w} \rp$
can be brought to the form
\be\label{eq:det_h1h0}
\det\lp 1+e^{\hat w} \rp = Z \det \lp 1- n(\epsilon) + e^{-i\hat h_1 \tau}e^{i\hat h_0 \tau} n(\epsilon)\rp
\ee
The operator $e^{-i\hat h_1 \tau}e^{i\hat h_0 \tau}$ is represented most 
naturally in the basis of time-dependent scattering states
constructed as wavepackets labeled by the time of arrival 
at the scatterer. As Fig.\,\ref{fig2} illustrates, 
the result of backward-and-forward time evolution is
\be
\hat S \equiv e^{-i\hat h_1 \tau}e^{i\hat h_0 \tau}=
\delta_{t,t'}\times \cases{ R,& $0<t<\tau$ \cr 1, & else }
\ee
with $R=S_1 S_0^{-1}$ a compound S-matrix, 
and $S_{0,1}$ the S-matrices 
for the charged/uncharged localized state. 
(In the single-channel case, $R=e^{2i(\delta-\delta')}$.) 
Thus we rewrite Eq.(\ref{eq:det_h1h0}) as
$
\det\lp 1+e^{\hat w} \rp =Z \det \lp 1+(\hat S-1) \hat n\rp
$.
Similarly, the operator $\lp 1+e^{-\hat w} \rp^{-1}$ becomes
\be\label{eq:inverse}
\lp 1+e^{-\hat w} \rp^{-1}= \lp n(\epsilon) + (1- n(\epsilon))\hat S^{-1}\rp^{-1} 
n(\epsilon)
\ee
with $n$, $S$ being operators
in the Hilbert space 
of functions of time. 

Thus the Green's function (\ref{eq:G_aa'})
is brought to the form 
\bea\label{eq:G_factored1}
&& G(\tau)=L e^{C}
\,,\quad
e^{C}=\det \lp 1+(\hat S-1) \hat n\rp
\\\label{eq:G_factored2}
&& L=\sum_{\epsilon,\epsilon'}
e^{-i\epsilon'\tau}
\la \tilde u_{\epsilon} |(1-\hat n)\lp \hat n+\hat S^{-1}(1- \hat n)\rp^{-1} | \tilde u^\ast_{\epsilon'}\ra
\eea
where $\tilde u_{\epsilon}=
\sum_\alpha u_\alpha \delta(\epsilon-\epsilon_\alpha)$
is a vector in channel space,
and 
$\la ...\ra$
includes summation over
scattering channels. (A related determinant identity for $e^C$ has been 
known in the theory of counting statistics~\cite{LL,Muzykantskii_Adamov}.)

The factorization $G(\tau)=L e^{C}$ 
demonstrated for a general scattering problem
with noncommuting $S_{0,1}$,
provides connection with
Nozieres-deDominicis theory and generalizes it to nonseparable scatterers.
The two factors in $G(\tau)$, expressed
through single-particle quantities,  
in the language of Ref.\cite{Nozieres_deDominicis} correspond
to the contributions of the open line and closed loop diagrams, 
respectively.

\begin{figure}[t]
\centerline{
\begin{minipage}[t]{3.2in}
\vspace{0pt}
\centering
\includegraphics[width=3.2in]{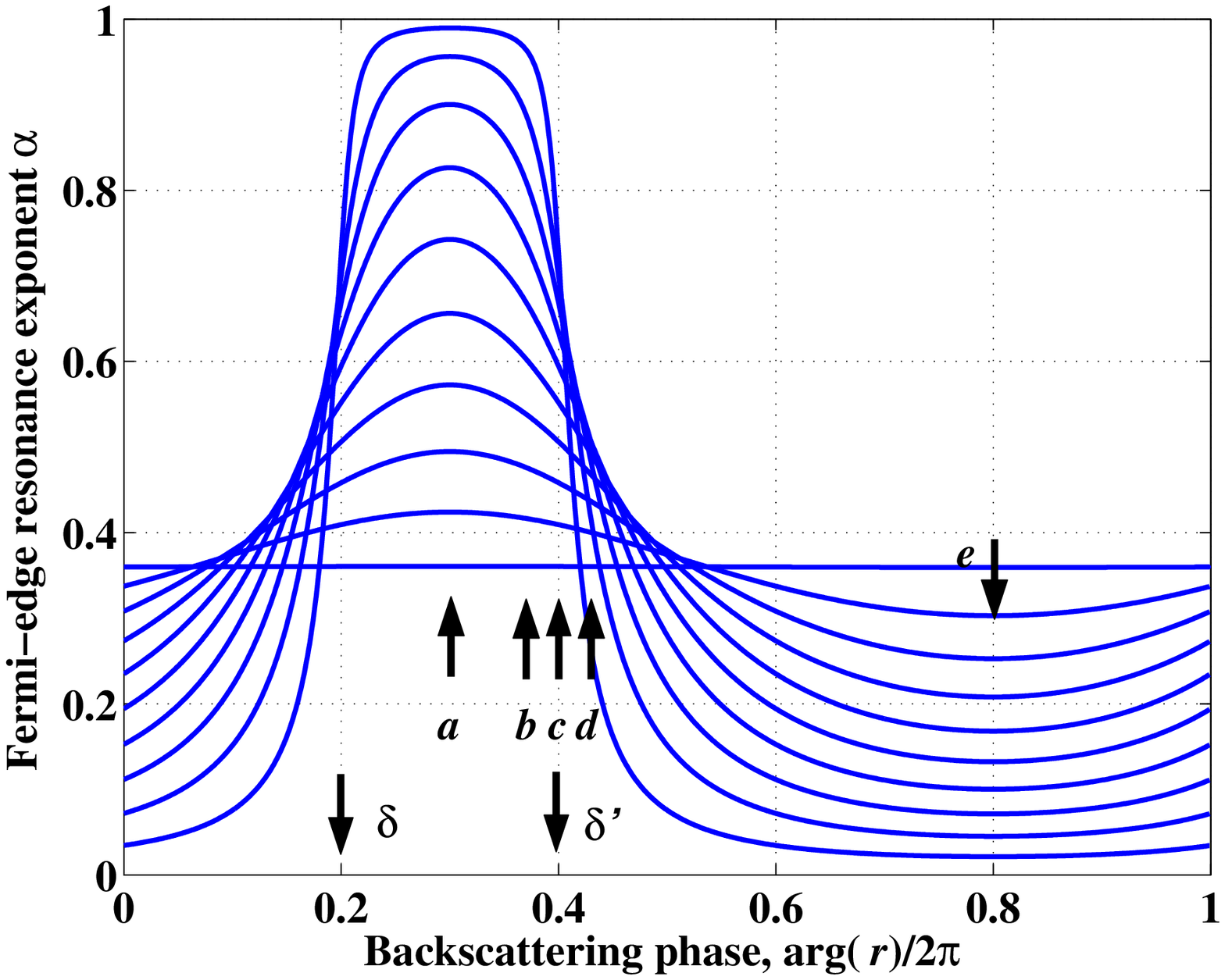}
\end{minipage}
\hspace{-1.7in}
\begin{minipage}[t]{1.65in}
\vspace{0.0in}
\centering 
\includegraphics[width=1.6in]{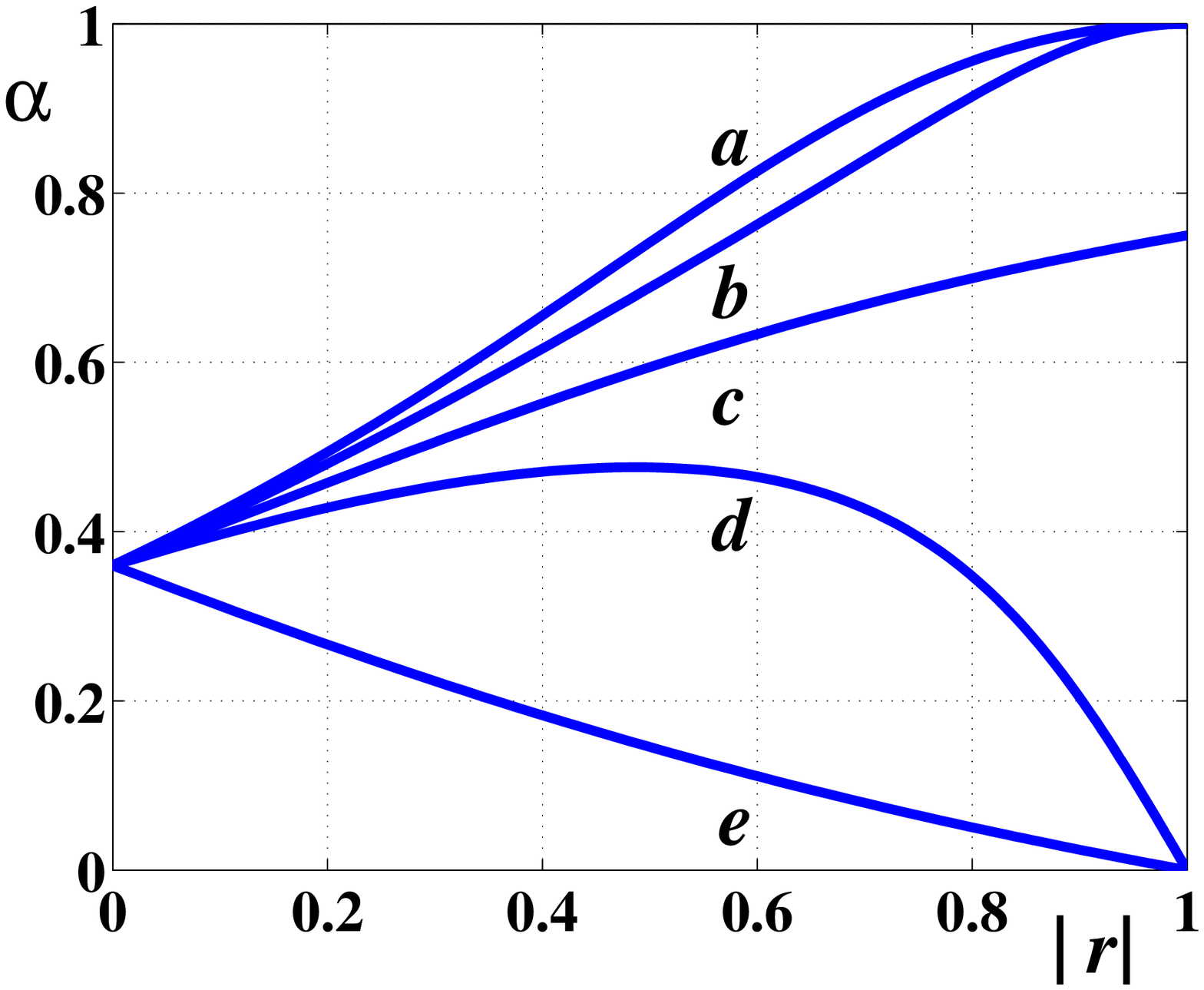}
\end{minipage}
}
\vspace{0.1cm}
\caption[]{
The dependence (\ref{eq:alpha_beta}),(\ref{eq:beta})
of the tunneling
exponent $\alpha$ on ${\rm arg}\,( r)$
for $|r|=0,\,0.1,...0.9$;
\emph{Inset:} The exponent $\alpha$ \emph{vs.} $|r|$
for several values 
${\rm arg}\, (r)$ 
marked by arrows (\emph{a---e}).
}
\label{fig3}
\end{figure}

An explicit result for $G(\tau)$ 
now follows by noting
that, with respect to channel indices, the operators in 
(\ref{eq:G_factored1}),(\ref{eq:G_factored2})
are diagonal in the eigenbasis of $R=S_1S_0^{-1}$,
where $L$ is additive, while $e^C$ is multiplicative. 
Using the standard 
singular integral 
equation solution\cite{Nozieres_deDominicis,YamadaYosida}, 
we obtain
$$
L=-\sum_j |u_j|^2 \frac{i}{\tau} (-i\tau\xi _0)^{2\beta_j}
,\ \ 
e^C=e^{-i\delta\mu t}(-i\tau\xi _0)^{-\sum_j\beta_j^2}  
$$
at $t < \hbar/T$, 
with $e^{2\pi i\beta_j }$ the eigenvalues of $R$,
and $\xi _0\simeq E_F$
the ultraviolet cutoff.
The prefactor $e^{-i\delta\mu t}$
describing the localized state energy offset
can be discarded.

In the case of our primary interest (\ref{eq:R}),
$R$ has only one nontrivial eigenvalue
$e^{2\pi i\beta}$, given by (\ref{eq:beta}),
and $u_i$ is an eigenvector $v_i$. This gives
$G(\tau)\propto \tau^{-(1-\beta)^2}$, leading directly to the result
(\ref{eq:alpha_beta}).
The dependence of the FES exponent $\alpha$ 
on $|r|$ and ${\rm arg}\, r$
is displayed in Fig.~\ref{fig3}.

The effect of mesoscopic fluctuations
on FES can be described by drawing $\tilde S$ from
a 
Wigner-Dyson ensemble of matrices 
of size $(N+1)\times (N+1)$, 
orthogonal, unitary or simplectic,
depending on the symmetry.
The backscattering amplitude $r$, 
being a diagonal matrix element of $\tilde S$, 
has a distribution\cite{BeenakkerRMP}
$P(r)\propto (1-|r|^2)^\gamma$ with $\gamma=N+1,\,(N+2)/2,\,2N+2$
for the three ensembles. This generates an FES exponent distribution of
width $\simeq \gamma^{-1/2}$ which is small
at large $N$. For fixed modulus $|r|$,
the change of the FES exponent
can be of either sign depending on the phase 
$\theta={\rm arg}\,r$ (Fig.~\ref{fig3}). The effect of scattering
is particularly prominent at $|r|$ approaching $1$,
where the FES is strongly enhanced for the phase values $\theta$ between 
$\delta$ and $\delta'$, and suppressed otherwise,
which corresponds to resonance formation in the dot.

In summary, this work presents an exact solution of the Fermi-edge resonance
problem for noncommuting scatterers, relevant for tunneling
in mesoscopic systems. We consider an application 
to resonant tunneling in open quantum dots
and show that 
a resonance with tunable interaction strength, 
and thus with a variable power law exponent, can be realized. 
The resonance is strongly enhanced by backscattering
in a phase-sensitive fashion.

This work has benefited from the discussions of
resonant tunneling spectroscopy of quantum dots
with Charles Marcus, Andrey Shytov and Dominik Zumb\"uhl,
and was  
supported by MRSEC Program of
the National Science Foundation
(DMR 02-13282).

\end{multicols}

\end{document}